\documentclass[letterpaper,11pt]{article}
\usepackage{amssymb,amsmath,epsf,euscript,graphicx,mathtools,authblk,setspace,enumerate,hyperref}
\usepackage{cite}

\newcommand{\beq}{\begin{equation}}
\newcommand{\eeq}{\end{equation}}
\newcommand{\barr}{\begin{array}}
\newcommand{\earr}{\end{array}}
\newcommand{\beqarr}{\begin{eqnarray}}
\newcommand{\eeqarr}{\end{eqnarray}}
\newcommand{\beqar}{\begin{eqnarray*}}
\newcommand{\eeqar}{\end{eqnarray*}}
\newcommand{\bef}{\begin{figure}}
\newcommand{\eef}{\end{figure}}

\newcommand{\bm}[1]{\mbox{\boldmath$#1$}}

\begin{document}

\title{The Effects of Wing Inertial Forces and Mean Stroke Angle on the Pitch Stability of Hovering Insects}
\author[1*]{Sevak Tahmasian}
\author[2]{Braeden C. Kotulak-Smith}
\affil[1]{Department of Biomedical Engineering and Mechanics, Virginia Tech Blacksburg, VA, USA, e-mail: sevakt@vt.edu}
\affil[2]{Kevin T. Crofton Department of Aerospace and Ocean Engineering, Virginia Tech Blacksburg, VA, USA}

\date{}
\maketitle
%\doublespacing
%%%%%%%%%%%%%%%%%%%%%%%%%%   Abstract   %%%%%%%%%%%%%%%%%%%%%%%%%%%%%%%%%%%
\begin{abstract}
This paper discusses the wing inertial effects on stability of pitch motion of hovering insects. The paper also presents a dynamic model appropriate for using averaging techniques and discusses the pitch stability results derived from the model. The model is used to predict the body angle of five insect species during hover, which are in good agreement with the available experimental results from different literature. The results suggest that the wing inertia forces have a considerable effect on pitch dynamics of insect flight and should not be ignored in dynamic analysis of hovering insects. The results also suggest that, though the pitch stability of hovering insects is open-loop stable, it may not be vibrationally stabilized. Instead, the pitch stability is a balance of the moment of insect's weight and the aerodynamic moment due to flapping kinematics with a nonzero mean stroke angle. Experiments with a flapping wing device confirm this results. To clearly explain the used model and clarify the difference between vibrational and non-vibrational stabilization, first this paper discusses the vibrational control of a three-degree-of-freedom force-input pendulum with its pivot moving in a vertical plane.
\end{abstract}

%%%%%%%%%%%%%%%%%%%%%%%%%%   Keywords   %%%%%%%%%%%%%%%%%%%%%%%%%%%%%%%%%%%
%\textbf{Keywords:} Vibrational control, Averaging, insect wing inertial forces, Kapitza pendulum, Pitch stability of hovering insects
%%%%%%%%%%%%%%%%%%%%%%%%%%   Section   %%%%%%%%%%%%%%%%%%%%%%%%%%%%%%%%%%%%
\section{Introduction}
\label{sec:Introduction}
%%%%%%%%%%%%%%%%%%%%%%%%%%%%%%%%%%%%%%%%%%%%%%%%%%%%%%%%%%%%%%%%%%%%%%%%%%%
Since Stephenson's observation that an inverted pendulum can be stabilized in its upright orientation by fast vertical vibrations of its pivot in the early twentieth century \cite{Stephenson.1908}, and the theoretical explanation of that phenomenon by Kapitza in the mid-twentieth century \cite{Kapitza.1965}, the Stephenson-Kapitza pendulum, usually called the Kapitza pendulum, has been the classical example of vibrational control and vibrational mechanics \cite{Meerkov.TAC.1980,Blekhman.2000}. The dynamics, stability analysis, and the mechanics underlying the stability of the Kapitza pendulum are discussed in different literature, for example \cite{Pippard.1987,Butikov.2001,Butikov.2011,Grundy.2019,Artstein.2021,Tahmasian&Woolsey.ND.2022}.

Developed by Meerkov, vibrational control is changing the stability properties of a dynamical system by introducing high-frequency, \emph{zero-mean} inputs to the system \cite{Meerkov.JFI.1977,Meerkov.TAC.1980}. The Kapitza pendulum benefits from the stabilizing effects of high-frequency, zero-mean, periodic inputs on mechanical systems for its stability. A well-developed, useful method for the dynamic analysis of mechanical systems with high-frequency periodic inputs is averaging. Using the averaging techniques, a time-periodic dynamical system can be approximated by a time-invariant system, called the averaged dynamics. For ``high enough'' frequencies, the existence of an asymptotically stable equilibrium point of the averaged dynamics guarantees the existence of an asymptotically stable periodic orbit of the time-periodic system in a small neighborhood of that equilibrium point  \cite{Guckenheimer&Holmes.1983,Sanders&Verhulst.1985}. Using the chronological calculus developed in \cite{Agrachev&Gamkrelidze.Sbornik.1978} and a series expansion that describes the evolution of mechanical systems subject to time-varying inputs \cite{Bullo.JCO.2001}, Bullo developed a closed form for the averaged dynamics of a class of control-affine mechanical systems \cite{Bullo.JCO.2002,Bullo&Lewis.2005}. The inclined Kapitza pendulum discussed in this paper belongs to this class of systems. Therefore, this paper uses the mentioned closed form averaging formula for stability analysis and vibrational control of the force-input inclined Kapitza pendulum and for stability analysis of the pitch motion of hovering insects also.

Insect flight has been the inspiration for the development of biomimetic flapping wing vehicles. Design of applicable and efficient flapping wing devices is directly related to our understanding of different aspects of insect flight. The longitudinal motion and pitch stability of hovering insects and flapping wing micro-air vehicles (FWMAVs) are vastly studied in different literature, with the analyses sometimes being inconclusive or the results being in contrast to previous ones. The analyses are all based on the aerodynamic forces and moments acting on the hovering insect or FWMAV. More often, the mass of the wings, wings inertial forces, the distance between the body center of mass and wing joints (wing hinge or root), and the asymmetry of the stroke (flapping) angle during one flapping cycle are neglected in the analyses \cite{Taylor&Thomas.JTB.2002,Faruque&HumbertP1.JTB.2010,Cheng&Deng.TR.2011,ChengDeng&Hedrick.JEB.2011,Karasek&Preumont.IJMAV.2012,Ristroph.JRSI.2013,ElzingaBreugel&Dickinson.BB.2014,TahaHajj&Nayfeh.JGCD.2014,TahaNayfeh&Hajj.ND.2014,TahaEtAl.BB.2015,Yao&Yeo.BB.2019,TahaEtAl.SR20,Saetti&Rogers.JGCD22}. Though in some research the wing inertial forces are also considered in the dynamics, the emphasis in those efforts are on the effects of the considered aerodynamic model and approximation methods on the pitch stability and control of insect or FWMAV flight, which, considering the role of aerodynamic forces in flight, is completely justified \cite{WuEtAl.JEB.2009,Orlowski&Girard.JGCD.2012,Wu&Sun.JRSI.2012,Sun.RMP.2014,KimEtAl.BB.2015,TahaWoolsey&Hajj.JGCD.2016,Bluman&Kang.BB.2017}. Recently the effects of the wing-to-body mass ratio and the body size of hovering insects on their power consumption are studied in \cite{XuEtAl.PF.2021,Lyu&Sun.JIP.2021}. However, so far, the vibrational stabilizing role of the wings inertial forces in the pitch stability of hovering flight has not been studied separately, and often, is completely neglected. The results of this study suggest that, depending on insect species, the time-periodic inertial forces can provide up to around $30\%$ of the necessary force for vibrationally stabilization of the pitch motion of hovering insects. However, these considerable inertial forces are neglected in the available research on vibrational stabilization of the pitch motion of hovering insects \cite{TahaEtAl.SR20}.

In Section~\ref{sec:Insect} of this paper, the effects of the wing inertial forces due to \emph{symmetric flapping} of the wings on the pitch stability of hovering insects are studied. It is shown that during hover, those inertial forces are comparable and most often larger than the aerodynamic forces acting on the body. However, though considerable in amplitude, they are not large enough to vibrationally stabilize the hovering insect body in a non-vertical orientation. Even the wing inertial forces and the aerodynamic forces and moments due to \emph{symmetric} flapping together are not large enough to \emph{vibrationally} stabilize the the insect body in a non-vertical orientation. Therefore, contrary to the results of \cite{TahaEtAl.SR20} (see also \cite{Karasek.SR.2020}), the results of this paper suggest that the pitch motion of hovering insects may not be \emph{vibrationally} stabilized after all.

Using the averaged dynamics of longitudinal motion of insect body, in Section~\ref{sec:InsectAF} of this paper it is shown that what may stabilize the pitch motion of an insect body during hovering flight is the \emph{nonzero-mean} aerodynamic moment due to the \emph{asymmetric} flapping. This result is also mentioned in \cite[Sec. 3.3]{EllingtonP3.PTRS.1984} without any analytical work. The body and wing damping causes the averaged pitch dynamics of the body to be asymptotically stable. In other words, the results suggest that the pitch motion of the hovering insects is open-loop stable, though not \emph{vibrationally} stable. The important effect of an asymmetric flapping kinematics on the pitch stability of hovering insects is also verified experimentally in the experiments discussed in this paper and presented in the accompanying video.

Using a simple aerodynamic model and physical and morphological parameters of five insect species, and using the averaged dynamics of longitudinal flight of insect body, the equilibrium orientation (body angle) of those species during hover is determined. Despite some assumptions, the numerical results show good agreement between the calculated body angles during hover and the body angles observed in experiments for the five insect species. The results also suggest that more important than the wing inertial forces, the flapping asymmetry (the nonzero-mean stroke angle) and the distance between the body center of mass and wing hinges constitute the main part of insect flight dynamics and should not be ignored in dynamic analysis of insect flight. The mean stroke angle and the distance between the body center of mass and wing hinges are usually considered ``small'' and neglected, for example, in \cite{TahaEtAl.SR20}. However, when talking about insects, those ``small'' parameters determine the major part of the pitch dynamics of hovering insects.

The entire work is based on first-order averaging techniques from \cite{Bullo.JCO.2002,Bullo&Lewis.2005,Vela&Morgansen&Burdick.ACC02} which proves to be sufficient for the stability analysis of hovering insects if observing all the conditions of the averaging theorem and the averaging technique presented in \cite{Bullo&Lewis.2005}.

The contributions of this paper are i) modeling of the pitch dynamics of hovering insects based on the dynamics of a 3-DOF, force-input inclined Kapitza pendulum in a form appropriate for averaging and controllability analysis while considering the wing inertial forces, ii) showing analytically, and experimentally verifying, that the pitch stability of hovering insects may not be vibrational, but can be due to a flapping kinematics with nonzero mean stroke angle, and iii) predicting the body angle of five insect species during hover with acceptable accuracy, which to the best of the authors' knowledge, is done for the first time. Without a doubt, a deeper understanding of the flight of insects will be helpful in developing more efficient and practical biomimetic air vehicles.

This paper is organized as follows. Section~\ref{sec:CIKP} discusses the dynamics and vibrational stabilization of a 3-DOF force-input pendulum. The dynamics and pitch stability of hovering insects are presented in Section~\ref{sec:Insect}. Section~\ref{sec:Conclusion} briefly discusses the main results of the paper.

%%%%%%%%%%%%%%%%%%%%%%%%%%%   Section   %%%%%%%%%%%%%%%%%%%%%%%%%
\section{The 3-DOF Force-Input Pendulum}
\label{sec:CIKP}
\begin{figure}[thpb]
\centering
\includegraphics[width=3.5 in]{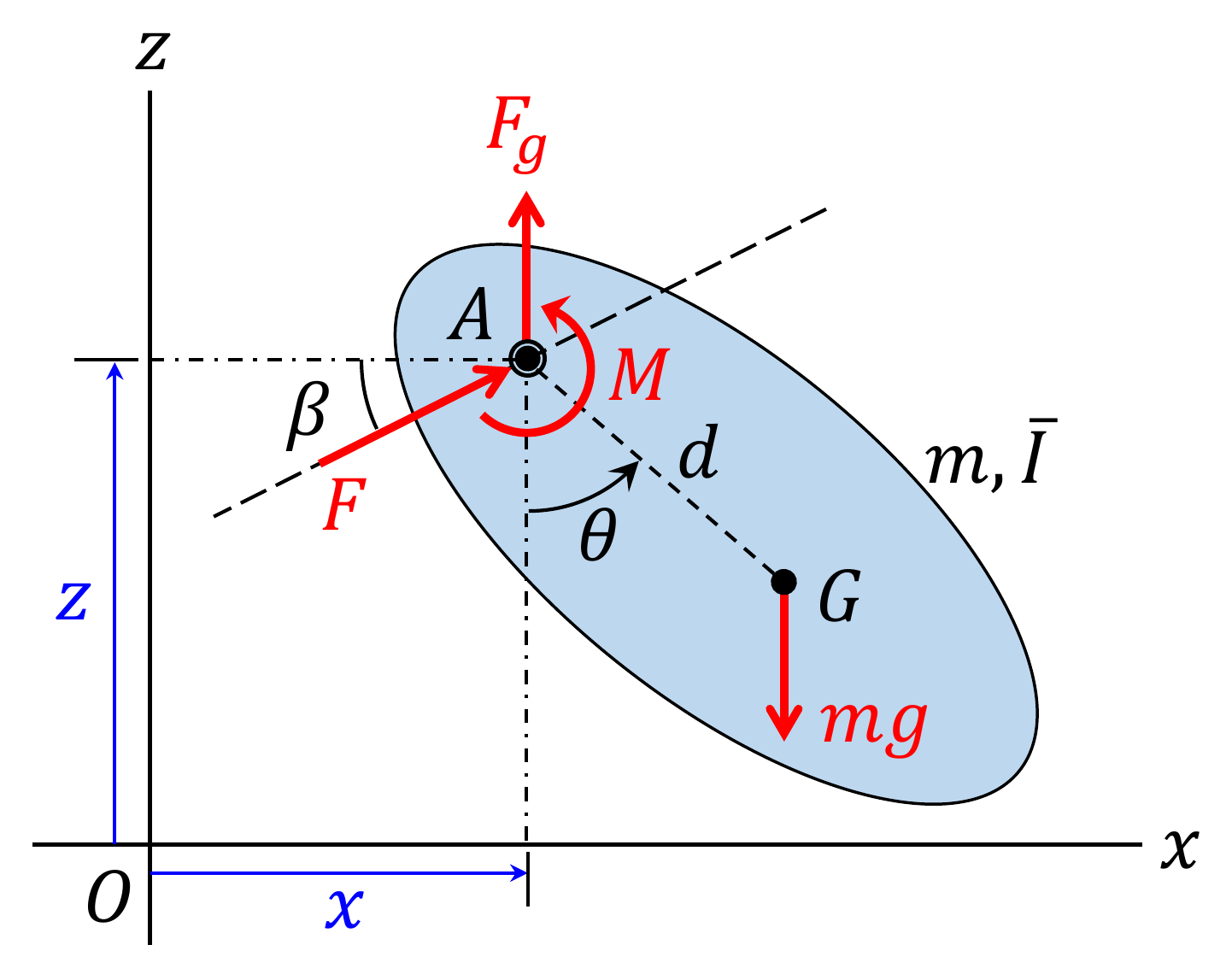}
\caption{The 3-DOF pendulum.}
\label{fig:CP}
\end{figure}

%%%%%%%%%%%%%%%%%%%%%%%%%%%%%%%%%%%%%%%%%%%%%%%%%%%%%%%%%%%%%%%%%%%%
Consider the 3-DOF pendulum (rigid body) of mass $m$ and mass moment of inertia $\bar{I}$ about its center of mass $G$ depicted in Figure~\ref{fig:CP}. The orientation angle $\theta$ of the pendulum is measured from its down position, that is, the $-z$-axis. The pivot $A$ of the pendulum, located at a distance $d$ from $G$, moves in the vertical $x$-$z$ plane under action of the input force $F$ applied in a fixed direction of angle $\beta$ with the horizontal $x$-axis. A constant force, $F_g= m g$, is applied to the pivot in the vertical direction, counteracting the weight of the pendulum. Therefore, the pendulum does not experience any weight, although the moment of the weight about the pivot is still present. An input couple $M$ also acts on the pendulum, as shown in Figure~\ref{fig:CP}. Small linear dampings with damping coefficients $c$ and $c_t$ resist the translational motion and rotation of the body, respectively.

%%%%%%%%%%%%%%%%%%%%%%%%%%%  FIGURE  %%%%%%%%%%%%%%%%%%%%%%%%%%%%%%%
%%%%%%%%%%%%%%%%%%%%%%%%%%%%%%%%%%%%%%%%%%%%%%%%%%%%%%%%%%%%%%%%%%%%
The equations of motion of the system are
\beqarr
m \ddot{x} + m d \ddot{\theta} \cos\theta - m d \dot{\theta}^2\sin\theta + c \dot{x} &=& F \cos\beta  \nonumber \\
m \ddot{z} + m d \ddot{\theta} \sin\theta + m d \dot{\theta}^2\cos\theta + c \dot{z} &=& F \sin\beta
\label{eqn:CPEOM} \\
m d \ddot{x}\cos\theta + m d \ddot{z}\sin\theta + I_A \ddot{\theta} + c_t \dot{\theta} + m g d\sin\theta &=& M  \nonumber
\eeqarr
where $x$ and $z$ are the coordinates of the position of the pivot, $I_A = \bar{I} + m d^2$ is the mass moment of inertia of the body about its pivot $A$. Consider the high-frequency, high-amplitude input force and couple in the form
\beqarr
F &=& F_0 \omega \varphi_1(\omega t) \nonumber \\
M &=& M_0 \omega \varphi_2(\omega t)
\label{eqn:CPForce}
\eeqarr
where $\varphi_1(t)$ and $\varphi_2(t)$ are zero-mean, $T$-periodic functions and $F_0 \geq 0$ and $M_0 \geq 0$ are constants. Following \cite{Bullo&Lewis.2005,TahmasianEtAl.JVC.2016}, the averaged dynamics of the system is
\beqarr
\ddot{\bar{x}} &=& -c\left(\frac{1}{m}+\frac{d^2}{\bar{I}}\cos^2\bar{\theta}\right)\dot{\bar{x}} - \frac{cd^2}{2\bar{I}}\dot{\bar{z}}\sin 2\bar{\theta}+\frac{c_t d}{\bar{I}}\dot{\bar{\theta}} \cos\bar {\theta} + d \dot{\bar{\theta}}^2\sin\bar{\theta} + f(\bar{\theta})d\cos\bar{\theta} \nonumber \\
\ddot{\bar{z}} &=& - \frac{cd^2}{2\bar{I}}\dot{\bar{x}}\sin 2\bar{\theta} -c\left(\frac{1}{m}+ \frac{d^2}{\bar{I}}\sin^2\bar{\theta}\right)\dot{\bar{z}} +\frac{c_t d}{\bar{I}}\dot{\bar{\theta}} \sin\bar{\theta}-d\dot{\bar{\theta}}^2\cos\bar{\theta} + f(\bar{\theta})d \sin\bar{\theta}
\label{eqn:CPAvgEq} \\
\ddot{\bar{\theta}} &=& \frac{cd}{\bar{I}}\dot{\bar{x}}\cos\bar{\theta}+\frac{cd}{\bar{I}}\dot{\bar{z}}\sin\bar{\theta}-\frac{c_t}{\bar{I}}\dot{\bar{\theta}} - f(\bar{\theta}) \nonumber
\eeqarr
where
\beq
f(\bar{\theta}) = \frac{mgd}{\bar{I}}\sin\bar{\theta} - \frac{\mu_{11}F_0^2d^2}{\bar{I}^2} \sin 2(\bar{\theta}-\beta) + \frac{2\mu_{12}F_0 M_0 d}{\bar{I}^2} \sin(\bar{\theta}-\beta)
\eeq
and where $\mu_{11} \geq 0$ and $\mu_{12}$ are determined using the periodic input functions $\varphi_1(t)$ and $\varphi_2(t)$ (see equation~\eqref{eqn:muij}) \cite{Bullo&Lewis.2005,TahmasianEtAl.JVC.2016}. For a brief review of the averaging technique used here, see \cite[Sec. A.2]{Tahmasian&Woolsey.ND.2022}.

Using the averaged dynamics~\eqref{eqn:CPAvgEq}, it can be shown that the orientation $\bar{\theta}_{\rm e}$ can be an equilibrium of the system if and only if
\beq
\mu_{11}d F_0^2 \sin 2(\bar{\theta}_{\rm e}-\beta) -2\mu_{12} F_0 M_0 \sin(\bar{\theta}_{\rm e}-\beta) - mg\bar{I}\sin\bar{\theta}_{\rm e} = 0
\label{eqn:CPFM}
\eeq
Note that equation~\eqref{eqn:CPFM} does not guarantee the stability of the equilibrium. Using linearization of the averaged dynamics, the equilibrium $\bar{\theta}_{\rm e}$ is stable if the following inequality is satisfied
\beq
2\mu_{11}d F_0^2 \cos 2(\bar{\theta}_{\rm e}-\beta) -2\mu_{12} F_0 M_0 \cos(\bar{\theta}_{\rm e}-\beta) - mg\bar{I}\cos\bar{\theta}_{\rm e} < 0
\label{eqn:CPFMStability}
\eeq

Consider the pendulum without the couple $M$, that is, $M_0=0$. For this case the necessary force amplitude to stabilize the pendulum in an orientation $\bar{\theta}_{\rm e}$ in its stabilizable set \cite{Tahmasian&Woolsey.ND.2022} is
\beq
F_0 = \sqrt{\frac{m g \bar{I} \sin\bar{\theta}_{\rm e}}{\mu_{11} d \sin 2(\bar{\theta}_{\rm e}-\beta)}}
\label{eqn:CPF0}
\eeq
It is evident that the pendulum with no couple $M$ can be stabilized in any orientation $\{\beta < \theta < \frac{\pi}{2}+\beta\} \cup \{-\frac{\pi}{2}+\beta < \theta \leq 0\}$, besides a third region entirely in the upper half-plane. In this section and Section~\ref{sec:Insect}, only the equilibria of the pendulum in the lower half-plane are considered. 

For the case of a horizontal input force only, i.e., $\beta=0$ and $M_0=0$, the pendulum can be stabilized in any orientation in the lower half-plane, that is, $-\frac{\pi}{2} < \bar{\theta}_{\rm e} < \frac{\pi}{2}$ using a force amplitude 
\beq
F_0 = \sqrt{\frac{m g \bar{I}}{2\mu_{11} d \cos\bar{\theta}_{\rm e}}}
\label{eqn:CPHF0}
\eeq
Equation~\eqref{eqn:CPHF0} suggests that to stabilize the pendulum in a non-vertical orientation in the lower half-plane using only a horizontal input force, the required force amplitude $F_0$ must be greater than a minimum force amplitude $F_{\rm m}=\sqrt{\frac{m g \bar{I}}{2\mu_{11} d}}$. For any force amplitude $F_0 \leq F_{\rm m}$ the pendulum remains in the downright orientation $\theta=0$, on average. Note that if $F_0=0$, then equation~\eqref{eqn:CPFM} cannot be satisfied except for $\bar{\theta}_{\rm e}=0$ or $\bar{\theta}_{\rm e}=180^{\circ}$, that is, the downright and upright orientations. This means it is not possible to stabilize the pendulum in a non-vertical orientation using a \emph{zero-mean} couple $M$ only.

Using a zero-mean force $F$ and a zero-mean couple $M$, however, the pendulum can be stabilized in a non-vertical orientation with a smaller or larger force amplitude $F_0$ (depending on the force and couple zero-mean functions $\varphi_1(t)$ and $\varphi_2(t)$) compared to the system with no couple. For example, consider the harmonic force and couple functions $\varphi_1 = \cos(t)$ and $\varphi_2 = \cos(t+\psi)$. For these functions, one determines $\mu_{11} = \frac{1}{4}$ and $\mu_{12} = \frac{1}{4}\cos\psi$. Therefore, for example, for the case of a horizontal force ($\beta=0$), using~\eqref{eqn:CPFM}, the required force amplitude to stabilize the pendulum in an orientation $\bar{\theta}_{\rm e}$ is determined to be
\beq
F_0 = \frac{1}{2d\cos\bar{\theta}_{\rm e}} \left(M_0\cos\psi + \sqrt{M_0^2\cos^2\psi + 8mgd\bar{I}\cos\bar{\theta}_{\rm e}} \right)
\label{eqn:CPFphase}
\eeq
It is evident that for $\psi = \pm 90^{\circ}$ the couple $M$ does not have any effect on the required force amplitude $F_0$, for $\psi=0$ the force amplitude $F_0$ is maximum, and for $\psi=180^{\circ}$ the force amplitude $F_0$ is minimum. Therefore, to stabilize the pendulum in an orientation $\bar{\theta}_{\rm e}$ using a zero-mean horizontal force and a zero-mean couple, by choosing $\psi=180^{\circ}$, the task can be accomplished with a smaller force amplitude. 

For any value of $\psi$, in general, the minimum horizontal force amplitude is determined using $\bar{\theta}_{\rm e}=0$, and the result is
\beq
F_{\rm m} = \frac{1}{2d} \left(M_0\cos\psi + \sqrt{M_0^2\cos^2\psi + 8mgd\bar{I}} \right)
\label{eqn:Fm}
\eeq
It is noteworthy that, in general, the equilibrium set and the stabilizable set (see \cite{Tahmasian&Woolsey.ND.2022}) of the pendulum with both force and couple inputs is different from those of the pendulum with only force input and the sets depend on the physical parameters of the pendulum and inputs. Depending on the couple amplitude $M_0$ and the periodic functions $\varphi_1(t)$ and $\varphi_2(t)$, the input couple may cause the equilibrium and stabilizable sets of the pendulum to expand or shrink.

\begin{figure}[thpb]
\centering
\includegraphics[width=5 in]{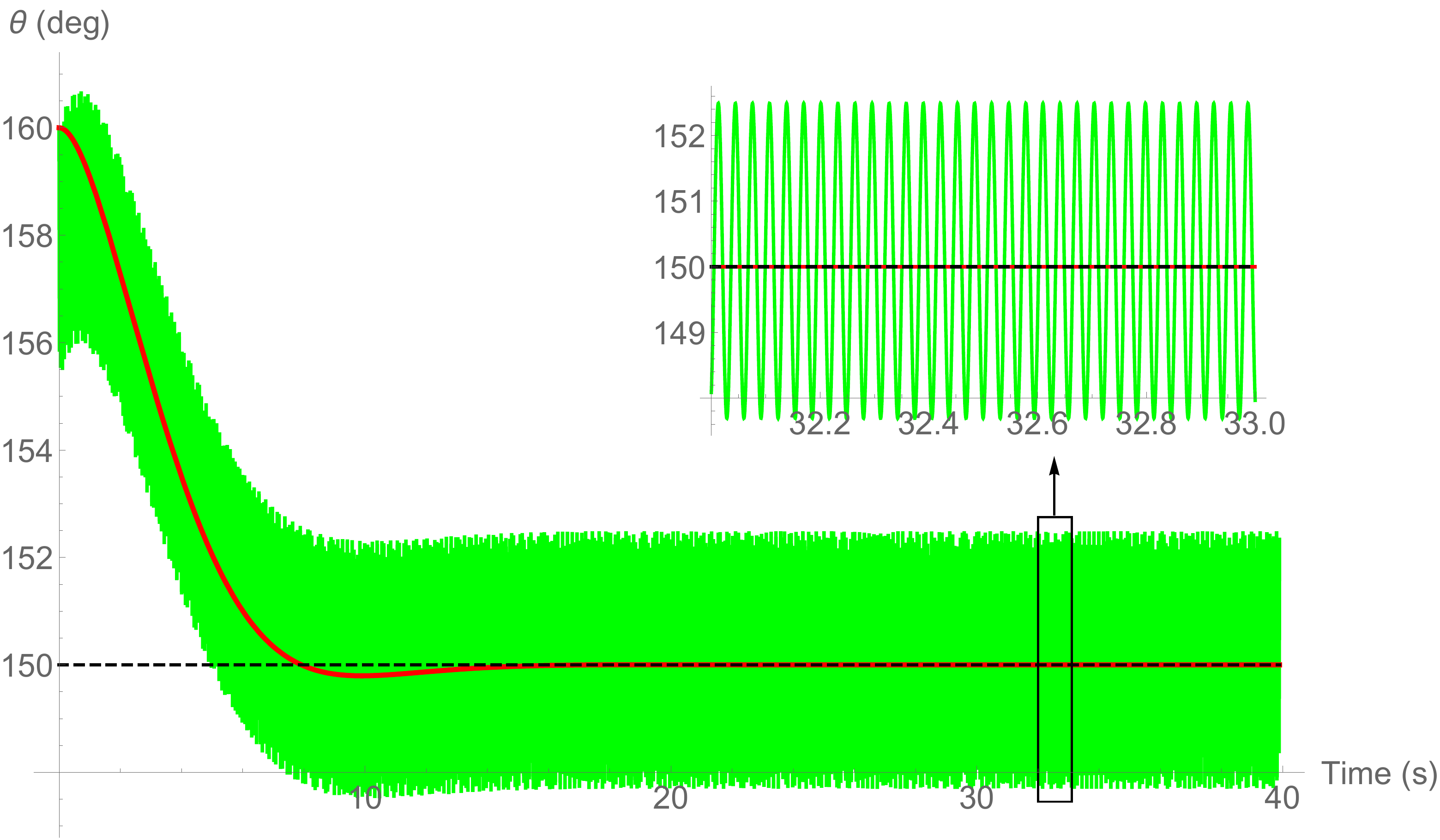}
\caption{Time history of the pendulum orientation stabilized at $\bar{\theta}_{\rm e}=150^{\circ}$ using a zero-mean horizontal force and a zero-mean couple. Solid-green: original system, solid-red: averaged system, dashed-black: desired orientation.}
\label{fig:CPFM}
\end{figure}

As mentioned, using a horizontal force only, the equilibrium and stabilizable sets of the 3-DOF pendulum are the lower half-plane. However, by adding a zero-mean couple, the pendulum can be stabilized in the upper half-plane as well. Figure~\ref{fig:CPFM} shows the time history of the pendulum stabilized at the desired orientation $\bar{\theta}_{\rm e}=150^{\circ}$ in the upper half-plane, on average, using a zero-mean horizontal force (i.e., $\beta=0$) and a zero-mean couple. The physical parameters are $m=0.2 \; {\rm kg}$, $d=0.2 \; {\rm m}$, $\bar{I}=0.05 \; {\rm kg.m^2}$, $c=0.1 \; {\rm N.s/m}$, $c_t=0.05 \; {\rm N.m.s/rad}$, $\omega=200 \; {\rm rad/s}$, $\varphi_1(t)=\cos t$, $\varphi_2(t)=-\cos t$ (and therefore, $\psi=180^{\circ})$, and $M_0=0.5 \; {\rm N.m}$. The initial conditions are $x(0)=z(0)=0$, $\theta(0)=160^{\circ}$, and zero initial velocities. From equation~\eqref{eqn:CPFM} one determines two values for the required force amplitude, $F_0=0.468 \; {\rm N}$ and $F_0=2.418 \; {\rm N}$. In the simulations the former is used. Note that without the input couple, the desired orientation $\bar{\theta}_{\rm e}=150^{\circ}$ will not belong to the equilibrium set of the system.

%%%%%%%%%%%%%%%%%%%%%%%%%%%  FIGURE  %%%%%%%%%%%%%%%%%%%%%%%%%%%%%%%
%%%%%%%%%%%%%%%%%%%%%%%%%%%%%%%%%%%%%%%%%%%%%%%%%%%%%%%%%%%%%%%%%%%%

%%%%%%%%%%%%%%%%%%%%%%%%%%%   Section   %%%%%%%%%%%%%%%%%%%%%%%%%%%%%%%%%%%
\section{Pitch Stability of Hovering Insects}
\label{sec:Insect}
%%%%%%%%%%%%%%%%%%%%%%%%%%%%%%%%%%%%%%%%%%%%%%%%%%%%%%%%%%%%%%%%%%%%%%%%%%%
This section discusses the effects of the wing inertial forces on vibrational stabilization of hovering insects and the dynamics and stability of the pitch motion of hovering insects. The body of a hovering insect can be considered as a 3-DOF pendulum, similar to what was discussed in Section~\ref{sec:CIKP}, with the aerodynamic forces and couples and inertial forces due to flapping of the wings acting on it. To discuss the effects of the wing inertial forces, in the first part of this section, all the aerodynamic forces and moments, except the average lift, are neglected and only the inertial forces due to the accelerating wings are considered as the input forces acting on the body. (The authors are aware that neglecting the aerodynamic forces in insect flight may seem surprising and is not justified for insect flight analysis. However, this is to discuss the effects of the wing inertial forces only. Besides, a considerable part of the aerodynamic forces, i.e., the averaged lift, is still considered.) In the second part, besides the wing inertial forces, an approximation of the aerodynamic forces and moments are also considered acting on the body and the pitch stability and body angle of five different insect species are determined.

%%%%%%%%%%%%%%%%%%%%%%%%%%%   Subsection   %%%%%%%%%%%%%%%%%%%%%%%%%%%%%%%%%%%
\subsection{The effect of wing inertial forces on pitch stability}
\label{sec:InsectIF}
\begin{figure}[thpb]
\centering
\includegraphics[width=5.8 in]{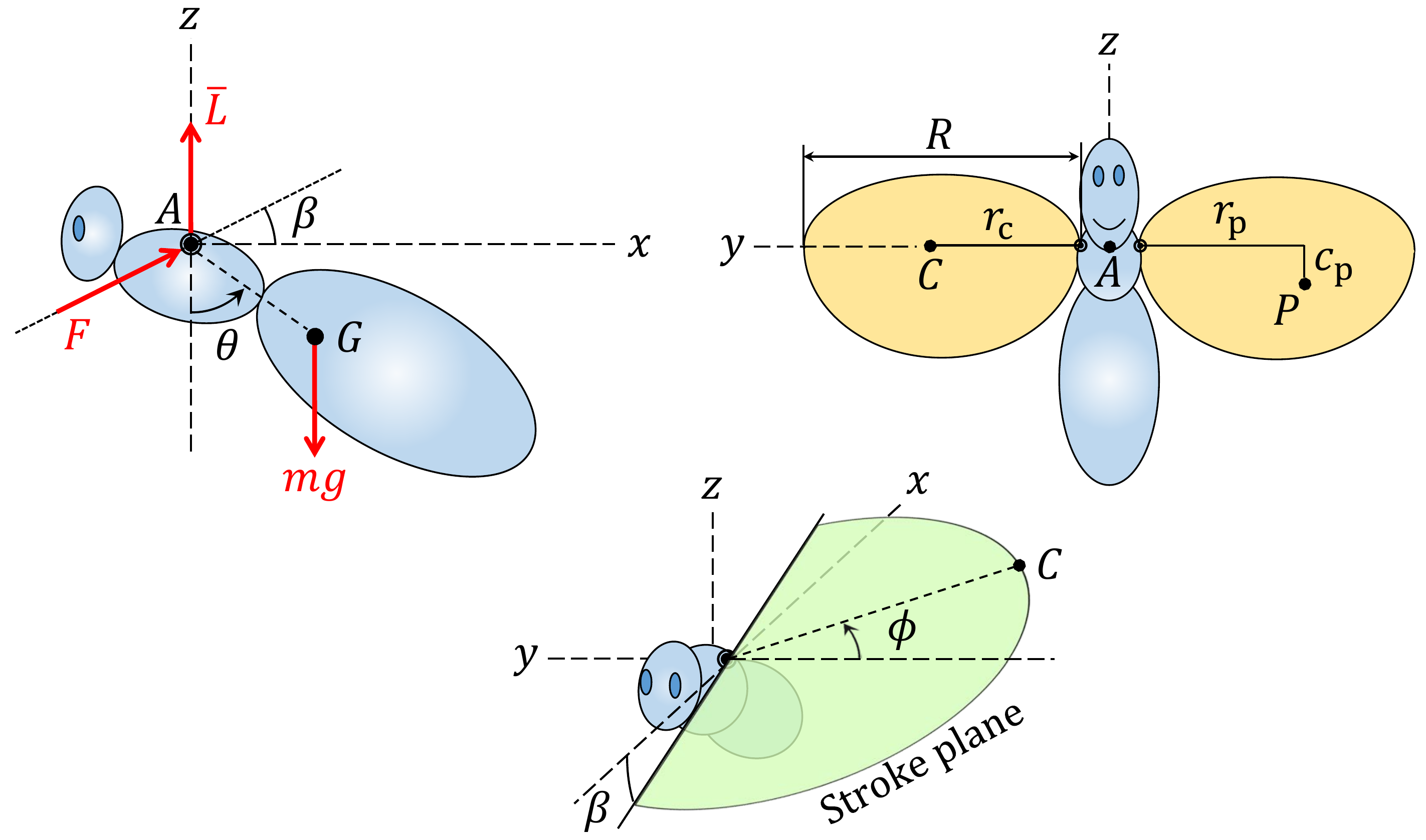}
\caption{The insect parameters and the inertial force $F$. The aerodynamic forces and moments, except the average lift $\bar{L}$, are not shown. The points $C$ and $P$ are the center of mass and center of pressure of the wings, respectively.}
\label{fig:Insect}
\end{figure}

%%%%%%%%%%%%%%%%%%%%%%%%%%%%%%%%%%%%%%%%%%%%%%%%%%%%%%%%%%%%%%%%%%%%%%%%%%%
Consider the hovering insect depicted in Figure~\ref{fig:Insect} with a body mass $m$, mass moment of inertia $\bar{I}$ of the body about its center of mass $G$ which is located at a distance $d$ from the $y$-axis passing through the wing joints, and with the midpoint between the two wing joints at $A$ on the $y$-axis, and therefore $AG=d$. The pitch angle $\theta$ of the body is defined as the angle of the line $AG$ with the vertical. 

%%%%%%%%%%%%%%%%%%%%%%%%%%%  FIGURE  %%%%%%%%%%%%%%%%%%%%%%%%%%%%%%%
%%%%%%%%%%%%%%%%%%%%%%%%%%%%%%%%%%%%%%%%%%%%%%%%%%%%%%%%%%%%%%%%%%%%

Suppose that the wings, each of mass $m_w$, length $R$, and average chord length $\bar{c}$, perform a harmonic flapping in the form
\beq
\phi (\omega t) = \phi_0 \sin\omega t
\label{eqn:Flapping}
\eeq
where $\phi(t)$ is the flapping (stroke) angle measured from the horizontal $y$-axis, $\phi_0>0$ is the flapping (stroke) amplitude, and $\omega$ is the (usually high) flapping frequency. For simplicity, assume that when $\phi=0$, the center of mass of each wing is on the $y$-axis. Therefore, the center of mass $C$ of each wing moves in a plane, called the stroke plane, which passes through the wing hinge and has an angle $\beta$ with the horizontal (see Figure~\ref{fig:Insect}). The total inertial force acting on the body due to the accelerating wings is determined to be
\beq
F = 2m_w r_{\rm c} \phi_0 \omega^2 \varphi(\omega t)
\label{eqn:WingForce}
\eeq
where $r_{\rm c}$ is the distance between the wing joint $A$ and its center of mass $C$ (see Figure~\ref{fig:Insect}), and the zero-mean, $2\pi$-periodic function $\varphi(t)$ is
\beq
\varphi (t) = \sin t\cos \phi(t) +\phi_0\cos^2 t\sin \phi(t).
\label{eqn:varphi}
\eeq
The inertial force $F$ of the wings is in the form of~\eqref{eqn:CPForce} with $F_0=2m_w r_{\rm c} \phi_0 \omega$. The force $F$, which lies in the stroke plane, has a constant angle $\beta$ with the horizontal. Besides the wing inertial forces, the aerodynamic forces and couple are also applied to the body at point $A$. During hover, the vertical component of the aerodynamic forces, called lift ($L$) in this paper, is equal to the total weight of the insect, on average, i.e., $\bar{L} = m_t g$. Also, with the considered wing kinematics~\eqref{eqn:Flapping}, the horizontal component of the aerodynamic forces (the projection of the aerodynamic force in the $x$-$y$ plane), called drag ($D$) in this paper, and its $x$-component $D_x$, are zero on average, i.e., $\bar{D}=\bar{D}_x=0$. Since the lift and drag are in the same order of magnitude and during hover $\bar{L}=m_t g$, the aerodynamic forces during hover are in the order of the weight of the insect. 

Neglecting the effects of the aerodynamic forces, except the average lift $\bar{L}$, the equations of motion of the insect body are in the form
\beqarr
m_t \ddot{x} + m d \ddot{\theta} \cos\theta - m d \dot{\theta}^2\sin\theta + c \dot{x} &=& F \cos\beta  \nonumber \\
m_t \ddot{z} + m d \ddot{\theta} \sin\theta + m d \dot{\theta}^2\cos\theta + c \dot{z} &=& F \sin\beta
\label{eqn:InsectEOM1} \\
m d \ddot{x}\cos\theta + m d \ddot{z}\sin\theta + I_A \ddot{\theta} + c_t \dot{\theta} + m g d\sin\theta + \bar{M}_w &=& 0  \nonumber
\eeqarr
where $m_t=m+2m_w$ is the total mass of the insect, $I_A = \bar{I} + md^2$ is the mass moment of inertia of the body about the $y$-axis passing through the wing hinge $A$, and 
\beq
\bar{M}_w = 2m_w g r_{\rm c}\sin\bar{\phi}\cos\beta
\label{eqn:Mbw}
\eeq
is the mean moment due to the weight of the wings (which may be ignored due to being small). Equations~\eqref{eqn:InsectEOM1} are in the averaging form presented in \cite{Bullo.JCO.2002,Bullo&Lewis.2005}. Using the averaged dynamics, which are not presented here, the required inertial force amplitude $F_0$ for stabilizing the insect body in an orientation $\bar{\theta}_{\rm e}$ in its stabilizable set of the system is determined to be
\beq
F_0 = 2m_w r_{\rm c} \phi_0\omega = \frac{1}{md}\sqrt{\frac{m_t (m_t I_A - m^2 d^2)(\bar{M}_w+mgd\sin\bar{\theta}_{\rm e})}{\mu \sin2(\bar{\theta}_{\rm e}-\beta)}}
\label{eqn:InsectF0}
\eeq
where the parameter $\mu>0$ is determined using the periodic function $\varphi(t)$ in~\eqref{eqn:varphi} \cite{TahmasianEtAl.JVC.2016}. For a certain value of the stroke plane angle $\beta$, the minimum value of $F_0$ to stabilize the body in a non-vertical orientation $0<\theta< 90^{\circ}+\beta$ is
\beq
F_m = \frac{1}{md}\sqrt{\frac{m_t (m_t I_A - m^2 d^2)(\bar{M}_w+mgd\sin\theta_{\rm m})}{\mu \sin2(\theta_{\rm m}-\beta)}}
\label{eqn:InsectFm}
\eeq
where $0 \leq \theta_{\rm m} <90^{\circ}$ is determined from the equation
\beq
\tan 2(\theta_{\rm m}-\beta) - 2\tan\theta_{\rm m} = \frac{2\bar{M}_w}{mgd\cos\theta_{\rm m}}
\eeq
The morphological properties of five insect species, namely, hawkmoth (HW), hoverfly (HF), dronefly (DF), honeybee (HB), and bumblebee (BB) used in this section are presented in Tables~\ref{tab:Table1} and~\ref{tab:Table2}. The data for each of the mentioned insect species are taken from  \cite{KimEtAl.BB.2015,MouEtAl.JEB.2011,Liu&Sun.JEB.2008,Ristroph.JRSI.2013,Dudley&Ellington.JEB.1990}, respectively. The underlined data could not be found in literature and are estimated. The values of $r_{\rm p}$ are considered around $60\%$-$70\%$ of the wing length \cite{DengEtAlP1.TRO.2006}. In Table~\ref{tab:Table3}, the total weight of the insect $W_t=m_t g$, the real amplitude of the total wing inertial forces determined using~\eqref{eqn:WingForce}, that is, $F_t=F_0 \omega = 2 m_w r_{\rm c} \phi_0 \omega^2$, the ratio $\rho_t=\frac{F_t}{W_t}$, and the input parameter $\mu$ determined using the periodic function $\varphi(t)$ in~\eqref{eqn:varphi} for the five insect species are presented. The ratio $\rho_0=\frac{F_0}{F_m}$, where $F_0=2m_w r_{\rm c} \phi_0\omega$ and $F_{\rm m}$ is the minimum force amplitude determined using~\eqref{eqn:InsectFm}, is also presented in Table~\ref{tab:Table3}. 

%%%%%%%%%%%%%%%%%%%%%%%%%%%%%%%   TABLE   %%%%%%%%%%%%%%%%%%%%%%%%%%%%%%%%%
\begin{table}[ht]
    \caption{The morphological properties of the body and flapping frequency of the five insect species. $m$: body mass, $\bar{I}$: mass moment of inertia about the center of mass, $d$: the distance from the center of mass to the $y$-axis, $\omega$: flapping frequency.}
    \begin{center}
    \begin{tabular}{|c|c|c|c|c|}
    \cline{2-5}
    \multicolumn{1}{c|}{} & \rule{0pt}{12pt} $m \: ({\rm mg})$ & $\bar{I} \: ({\rm mg.cm^2})$ & $d \: ({\rm mm})$ & $\omega \: ({\rm Hz})$ \\
    \hline
    HM & 1360 & 2830 & 10.9 & 28.4 \\

    HF & 10.4 & 0.51 & 0.88 & 164 \\
    
    DF & 88.9 & 11.8 & 1.84 & 164 \\
    
    HB & 102 & 22.0 & 3.30 & 197 \\
    
    BB & 175 & 21.3 & 3.91 & 155 \\
    \hline
    \end{tabular}
    \label{tab:Table1}
\end{center}
\end{table}
%%%%%%%%%%%%%%%%%%%%%%%%%%%%%%%%%%%%%%%%%%%%%%%%%%%%%%%%%%%%%%%%%%%%%%%%%%%

%%%%%%%%%%%%%%%%%%%%%%%%%%%%%%%   TABLE   %%%%%%%%%%%%%%%%%%%%%%%%%%%%%%%%%
\begin{table}[ht]
    \caption{The morphological properties of the wings and flapping kinematics parameters of the five insect species. $m_w$: mass of one wing, $R$: wing length, $\bar{c}$: mean chord length, $r_c$: distance from the wing hinge to wing center of mass, $r_{\rm p}$: distance from wing hinge to center of pressure in $y$-direction, $c_{\rm p}$ distance from center of pressure to the $y$-axis, $\phi_0$: stroke amplitude, $\bar{\phi}$: mean stroke angle, $\beta$: angle of stroke plane.}
    \begin{center}
    \begin{tabular}{|c|c|c|c|c|c|c|c|c|c|}
    \cline{2-10}
    \multicolumn{1}{c|}{} & \rule{0pt}{12pt} $m_w \: ({\rm mg})$ & $R \: ({\rm mm})$ & $\bar{c} \: ({\rm mm})$ & $r_{\rm c} \: ({\rm mm})$ & $r_{\rm p} \: ({\rm mm})$ & $c_{\rm p} \: ({\rm mm})$ & $\phi_0 \: (^{\circ})$ & $\bar{\phi} \: (^{\circ})$ & $\beta \: (^{\circ})$ \\
    \hline
    HM & 48.3 & 48 & 18.1 & \underline{20} & \underline{30} & \underline{4.5} & 60 & 15 & 10 \\

    HF & 0.20 & 7.1 & 1.7 & \underline{3.5} & \underline{4.4} & \underline{0.4} & 42 & 12 & 25 \\
    
    DF & 0.56 & 11.2 & 3 & \underline{5.6} & \underline{7.0} & \underline{0.75} & 54 & 7 & 4 \\
    
    HB & 0.26 & 9.8 & \underline{2.5} & \underline{5.0} & \underline{7.0} & \underline{0.6} & 66 & 18 & \underline{0} \\
    
    BB & 0.46 & 13.2 & 4.0 & \underline{6.5} & \underline{7.5} & \underline{1.0} & 58 & 20 & 6 \\
    \hline
    \end{tabular}
    \label{tab:Table2}
\end{center}
\end{table}
%%%%%%%%%%%%%%%%%%%%%%%%%%%%%%%%%%%%%%%%%%%%%%%%%%%%%%%%%%%%%%%%%%%%%%%%%%%

%%%%%%%%%%%%%%%%%%%%%%%%%%%%%%%   TABLE   %%%%%%%%%%%%%%%%%%%%%%%%%%%%%%%%%
\begin{table}[ht]
    \caption{The weight and inertial forces of the five insect species.}
    \begin{center}
    \begin{tabular}{|c|c|c|c|c|c|c|}
    \cline{2-7}
    \multicolumn{1}{c|}{} & \rule{0pt}{12pt} $W_t \: ({\rm mN})$ & $F_t \: ({\rm mN})$ & $\rho_t$ & $\mu$ & $\theta_{\rm m} \: (^{\circ})$ & $\rho_0$ \\
    \hline
    HM & 14.3 & 64.4 & 4.51 & 0.19 & 40.4 & 0.28 \\

    HF & 0.11 & 1.06 & 10.0 & 0.22 & 63.1 & 0.20 \\
    
    DF & 0.88 & 6.28 & 7.11 & 0.20 & 26.8 & 0.14 \\
    
    HB & 1.01 & 4.59 & 4.56 & 0.18 & 7.6 & 0.08 \\
    
    BB & 1.73 & 5.74 & 3.33 & 0.20 & 31.4 & 0.10 \\
    \hline
    \end{tabular}
    \label{tab:Table3}
\end{center}
\end{table}
%%%%%%%%%%%%%%%%%%%%%%%%%%%%%%%%%%%%%%%%%%%%%%%%%%%%%%%%%%%%%%%%%%%%%%%%%%%
There are two obvious conclusions from the values presented in Table~\ref{tab:Table3}. First, since during hover the aerodynamic forces are in the order of the body weight, from the values of the ratio $\rho_t$ it is concluded that during hover, the amplitude of the wing inertial forces may be larger than the aerodynamic forces and should not be neglected in stability analysis of hovering flight. And second, the values of the ratio $\rho_0$ suggest that since for the five insect species $\rho_0 < 1$, the inertial forces, though considerable in magnitude, are not large enough to stabilize the body in a non-vertical orientation. They only provide around $10\%\:$-$\:30\%$ of the required vibrational force to put the body in a non-vertical orientation. It must be emphasised that, the results are based on a number of assumptions discussed earlier, such as harmonic flapping of the wings and neglecting the fluid added mass. From the numerical values in Table~\ref{tab:Table3}, it is also evident that, since the aerodynamic lift and drag are in the range of the weight, and therefore smaller than the wing inertial forces, adding the aerodynamic forces only (and not aerodynamic moments) does not have a considerable effect on the results and does not change the result that vibrational forces cannot stabilize the body in a non-vertical orientation.

Considering the results presented in Section~\ref{sec:CIKP} about the role of a zero-mean couple on the necessary force amplitude, one may think of the role of the aerodynamic couple on the pitch stability. The flapping kinematics~\eqref{eqn:Flapping} considered in this section, generates a zero-mean aerodynamic moment. Since the aerodynamic moment generated due to the aerodynamic forces is a function of the square of the wing velocity, i.e., $\dot{\phi}^2$, it does not have a phase difference with the centripetal acceleration, and its phase difference with the tangential acceleration $\ddot{\phi}$ is $90^{\circ}$. Therefore, the phase angle between the inertial forces and aerodynamic moment is in the range of zero and $90^{\circ}$, and based on the discussions in Section~\ref{sec:CIKP}, the zero-mean aerodynamic couple does not help reducing the necessary force amplitude for stabilization of the body in a non-vertical orientation. In other words, with the \emph{symmetric} flapping kinematics~\eqref{eqn:Flapping}, all the aerodynamic and inertial forces and moments together are not large enough to stabilize the body of a hovering insect in a non-vertical orientation. The results suggest that though in the real world the bodies of hovering insects are stabilized in a non-vertical orientation, they are not \emph{vibrationally} stabilized. After introducing the aerodynamic parameters in Section~\ref{sec:InsectAF}, the results of a numerical simulation will be presented which confirm this claim (see Figure~\ref{fig:HMSym}).

%%%%%%%%%%%%%%%%%%%%%%%%%%%   Subsection   %%%%%%%%%%%%%%%%%%%%%%%%%%%%%%%%%%%
\subsection{Hovering insects dynamics with aerodynamic and inertial forces}
\label{sec:InsectAF}
%%%%%%%%%%%%%%%%%%%%%%%%%%%%%%%%%%%%%%%%%%%%%%%%%%%%%%%%%%%%%%%%%%%%%%%%%%%
In this section, considering both the aerodynamic and inertial forces and moments acting on an insect body, its averaged dynamics are derived and the body angle during hover is predicted. The \emph{symmetric} flapping kinematics~\eqref{eqn:Flapping} is not the real kinematics that insects perform during hover. A more realistic kinematics is in the \emph{asymmetric} form \cite{EllingtonP3.PTRS.1984}
\beq
\phi(\omega t) = \bar{\phi} + \phi_0 \: \zeta(\omega t)
\label{eqn:FlapKin1}
\eeq
where $\bar{\phi}> 0$ is a constant, called the mean stroke angle, $\phi_0>0$ is the stroke amplitude, and $\zeta(t)$ is a zero-mean, $T$-periodic function. In this section the harmonic function $\zeta(t)=\sin t$ is used which is an acceptable estimation of flapping kinematics for most of the insects. Using the asymmetric kinematics~\eqref{eqn:FlapKin1}, insects generate a \emph{nonzero-mean} aerodynamic couple which opposes the moment of their weight about the wing hinges and stabilizes their body in a non-vertical orientation \cite{EllingtonP3.PTRS.1984}. Compared to~\eqref{eqn:InsectEOM1}, a more general dynamics of the insect body presented in Figure~\ref{fig:Insect} is
\beqarr
m_t \ddot{x} + m d \ddot{\theta} \cos\theta - m d \dot{\theta}^2\sin\theta + c \dot{x} &=& F_x  \nonumber \\
m_t \ddot{z} + m d \ddot{\theta} \sin\theta + m d \dot{\theta}^2\cos\theta + c \dot{z} + m_t g &=& F_z
\label{eqn:InsectEOM2} \\
m d \ddot{x}\cos\theta + m d \ddot{z}\sin\theta + I_A \ddot{\theta} + c_t \dot{\theta} + m g d\sin\theta + \bar{M}_w &=& M_y  \nonumber
\eeqarr
where $F_x$ and $F_z$ are the total (aerodynamic and inertial) forces in the $x$- and $z$-directions, $M_y$ is the total aerodynamic moment acting on the body about the $y$-axis, and $\bar{M}_w$ is defined in~\eqref{eqn:Mbw}.

Assuming a quasi-steady aerodynamic model, the aerodynamic forces are proportional to the square of the wing velocities, $\dot{\phi}^2$. Therefore, the lift force $L$ is
\beq
L = L_0 \dot{\phi}^2 = L_0 \phi_0^2 \omega^2 \cos^2\omega t
\eeq
where $L_0$ is a constant. The average lift $\bar{L}$ is determined being
\beq
\bar{L} = \frac{1}{T} \int_0^T L dt = \frac{1}{2} L_0 \phi_0^2 \omega^2
\eeq
Since during hover $\bar{L}=m_t g$, one determines $L_0 = \frac{2m_tg}{\phi_0^2\omega^2}$, and the lift force can be written in the form
\beq
L = 2m_t g\cos^2\omega t = m_tg(1+\cos2\omega t)
\label{eqn:Lift}
\eeq
During hover the drag $D$ is a zero-mean force which is also proportional to $\dot{\phi}^2$. For many insects, the lift-to-drag amplitude ratio during hover is almost one. Therefore, this paper assumes that the amplitude of the drag is equal to the amplitude of the lift, that is,
\beq
D = m_t g(1+\cos2\omega t) {\rm sgn}(\dot{\phi}) = m_t g(1+\cos2\omega t) {\rm sgn}(\cos\omega t)
\eeq
where ${\rm sgn}(\cdot)$ is the signum function. Therefore, the $x$ component of the drag is
\beq
D_x = -m_t g (1+\cos2\omega t)\cos\phi(\omega t) \: {\rm sgn}(\cos\omega t)
\label{eqn:Drag}
\eeq
To determine the aerodynamic moment, suppose that the pressure center of the wing is located at point $P$, a distance $r_{\rm p}$ and $c_{\rm p}$ from the wing root, as shown in Figure~\ref{fig:Insect}. Since the lift and drag are assumed equal, the total aerodynamic moment acting on the body about the $y$-axis is determined to be
\beqarr
M_y &=& 2m_t g\Big( \big( c_{\rm p} \sin\alpha \cos\phi(\omega t) + r_{\rm p} \sin\phi(\omega t) \big) \big( \cos\beta + \sin\beta \cos\phi(\omega t) \: {\rm sgn}(\cos\omega t) \big) - \nonumber \\
&& c_{\rm p} \cos\alpha \big(\sin\beta - \cos\beta \cos\phi(\omega t) \: {\rm sgn}(\cos\omega t) \big) \Big) \cos^2\omega t
\label{eqn:Moment}
\eeqarr
where $\alpha = \alpha_0 \: {\rm sgn} (\dot{\phi}) = \alpha_0 \: {\rm sgn}(\cos \omega t)$ is the wing pitch angle measured from the vertical. For the five insect species considered in this paper, it is assumed that $\alpha_0 = 45^{\circ}$. Adding the wing inertial forces $F$ from~\eqref{eqn:WingForce}, the forces $F_x$ and $F_z$ are
\beqarr
F_x &=& D_x + F \cos\beta \nonumber \\
F_z &=& L + F \sin\beta 
\label{eqn:Forces}
\eeqarr
It is evident that $F_x$ is zero-mean, however, $F_z$ and $M_y$ are not. To transform the equations of motion~\eqref{eqn:InsectEOM2} into an appropriate form for averaging, the total forces and moment can be rewritten in the form
\beqarr
F_x &=& \omega \varphi_1(\omega t) \nonumber \\
F_z &=& m_t g + \omega \varphi_2(\omega t)
\label{eqn:TotalForces} \\
M_y &=& \bar{M} + \omega \varphi_3(\omega t) \nonumber
\eeqarr
where using $T=\frac{2\pi}{\omega}$,
\beq
\bar{M} = \frac{1}{T} \int_0^{T} M_y dt
\eeq
and where the \emph{zero-mean}, $2\pi$-periodic functions $\varphi_i(t)$, $i \in \{1,2,3\}$, are
\beqarr
\varphi_1 (t) &=& -\frac{2m_t g}{\omega} \cos^2t\cos\phi(t) \: {\rm sgn}(\cos t) + 2m_w r_{\rm c} \phi_0 \omega \cos\beta \big( \sin t \cos\phi(t) + \phi_0 \cos^2 t \sin\phi(t) \big) \nonumber \\
\varphi_2(t) &=& \frac{m_t g}{\omega}\cos2t + 2m_w r_{\rm c} \phi_0 \omega \sin\beta \big( \sin t \cos\phi(t) + \phi_0 \cos^2 t \sin\phi(t) \big)
\label{eqn:varphis} \\
\varphi_3(t) &=& \frac{1}{\omega} \left[ 2m_t g\Big( \big( c_{\rm p} \sin (\alpha_0 \: {\rm sgn}(\cos t)) \cos\phi(t) + r_{\rm p} \sin\phi(t) \big) \big( \cos\beta + \sin\beta \cos\phi(t) \: {\rm sgn}(\cos t) \big) - \right. \nonumber \\
&& \left. c_{\rm p} \big(\sin\beta - \cos\beta \cos\phi(t) \: {\rm sgn}(\cos t) \big)\cos(\alpha_0 \: {\rm sgn}(\cos t)) \Big) \cos^2 t -\bar{M} \right] \nonumber 
\eeqarr
Replacing the total forces and moment from~\eqref{eqn:TotalForces} into~\eqref{eqn:InsectEOM2}, the equations of motion of the insect body are
\beqarr
m_t \ddot{x} + m d \ddot{\theta} \cos\theta &=& m d \dot{\theta}^2\sin\theta - c \dot{x} +\omega \varphi_1(\omega t)  \nonumber \\
m_t \ddot{z} + m d \ddot{\theta} \sin\theta &=& -m d \dot{\theta}^2\cos\theta - c \dot{z} + \omega \varphi_2(\omega t)
\label{eqn:InsectEOM3} \\
m d \ddot{x}\cos\theta + m d \ddot{z}\sin\theta + I_A \ddot{\theta} &=& - c_t \dot{\theta} - m g d\sin\theta - \bar{M}_w +\bar{M} + \omega \varphi_3(\omega t)  \nonumber
\eeqarr
Equations~\eqref{eqn:InsectEOM3} are in the averaging form presented in \cite{Bullo.JCO.2002,Bullo&Lewis.2005,TahmasianEtAl.JVC.2016}. The determined averaged dynamics are
\beqarr
\ddot{\bar{x}} &=& -\frac{c}{m_t}\dot{\bar{x}} + \frac{c_t md}{B} \dot{\bar{\theta}} \cos\bar{\theta} + \frac{md}{m_t}\dot{\bar{\theta}}^2\sin\bar{\theta} - \frac{md}{B}(\bar{M}-mgd\sin\bar{\theta}-\bar{M}_w)\cos\bar{\theta} - \frac{md}{m_t} \Delta \cos\bar{\theta} \nonumber \\
\ddot{\bar{z}} &=& -\frac{c}{m_t}\dot{\bar{z}} - \frac{c_t md}{B} \dot{\bar{\theta}} \sin\bar{\theta} - \frac{md}{m_t}\dot{\bar{\theta}}^2 \cos\bar{\theta} - \frac{md}{B}(\bar{M}-mgd\sin\bar{\theta}-\bar{M}_w) \sin\bar{\theta} - \frac{md}{m_t} \Delta \sin\bar{\theta} 
\label{eqn:InsectAvgEq} \\
\ddot{\bar{\theta}} &=& -\frac{c_t m_t}{B}\dot{\bar{\theta}} + \frac{m_t}{B}(\bar{M}-mgd\sin\bar{\theta}-\bar{M}_w) + \Delta \nonumber
\eeqarr
where
\[
B=m_t I_A-m^2d^2
\]
and
\[
\Delta = \frac{cmd}{B}(\dot{\bar{x}}\cos\bar{\theta}+\dot{\bar{z}}\sin\bar{\theta}) + \frac{md}{B^2}\Big( md\big((\mu_{11}-\mu_{22})\sin2\bar{\theta} - 2\mu_{12}\cos2\bar{\theta} \big) -2m_t(\mu_{13}\sin\bar{\theta} - \mu_{23} \cos\bar{\theta} ) \Big)
\]
and where, following \cite{TahmasianEtAl.JVC.2016}, the parameters $\mu_{ij}$, $i,j = 1,2,3$, are determined using the functions $\varphi_i(t)$, $i=1,2,3$, in the form
\beq
\mu_{ij} = \frac{1}{2T} \int_0^T \left(\int_0^t \varphi_i(\tau) d\tau \right)\left(\int_0^t \varphi_j(\tau) d\tau \right) dt - \frac{1}{2T^2} \left(\int_0^T \int_0^t \varphi_i(\tau) d\tau dt \right)\left(\int_0^T \int_0^t \varphi_j(\tau) d\tau dt \right)
\label{eqn:muij}
\eeq
Using the averaged dynamics~\eqref{eqn:InsectAvgEq}, the equilibrium orientation of the body, on average, is determined from equation $\ddot{\bar{\theta}} =0$ when replacing the average velocities with zero, i.e., $\dot{\bar{x}}=\dot{\bar{z}}=0$ and $\dot{\bar{\theta}}=0$, which is in the form
\beq
m_t(\bar{M}-mgd\sin\bar{\theta}) + \frac{md}{B}\Big( md\big((\mu_{11}-\mu_{22})\sin2\bar{\theta} - 2\mu_{12}\cos2\bar{\theta} \big) -2m_t(\mu_{13}\sin\bar{\theta} - \mu_{23} \cos\bar{\theta} ) \Big) =0
\label{eqn:BodyAngle}
\eeq
One may also use the approximation $m_t \approx m$ to write the averaged dynamics~\eqref{eqn:InsectAvgEq} and the equilibrium determining equation~\eqref{eqn:BodyAngle} in slightly simpler forms. 

Using the state vector $\bar{\bm{y}}=(\bar{\theta},\dot{\bar{x}},\dot{\bar{z}},\dot{\bar{\theta}})^T$, the averaged dynamics~\eqref{eqn:InsectAvgEq} can be written as a first order system, which then can be linearized about the equilibrium point $\bar{\bm{y}}_{\rm e}=(\bar{\theta}_{\rm e},0,0,0)^T$, where $\bar{\theta}_{\rm e}$ is the equilibrium orientation determined from~\eqref{eqn:BodyAngle}. The linearized averaged dynamics shows that for each of the five insect species with their morphological properties presented in Tables~\ref{tab:Table1} and~\ref{tab:Table2}, the determined equilibrium is stable. Therefore the original system possesses a stable periodic orbit in a small neighborhood of that equilibrium. The damping coefficients and the eigenvalues of the state matrix of the linearized averaged dynamics for the five insect species are presented in Table~\ref{tab:Table4}. The eigenvalues show the open-loop stability of the pitch motion for those species.

\begin{figure}[thpb]
\centering
\includegraphics[width=6 in]{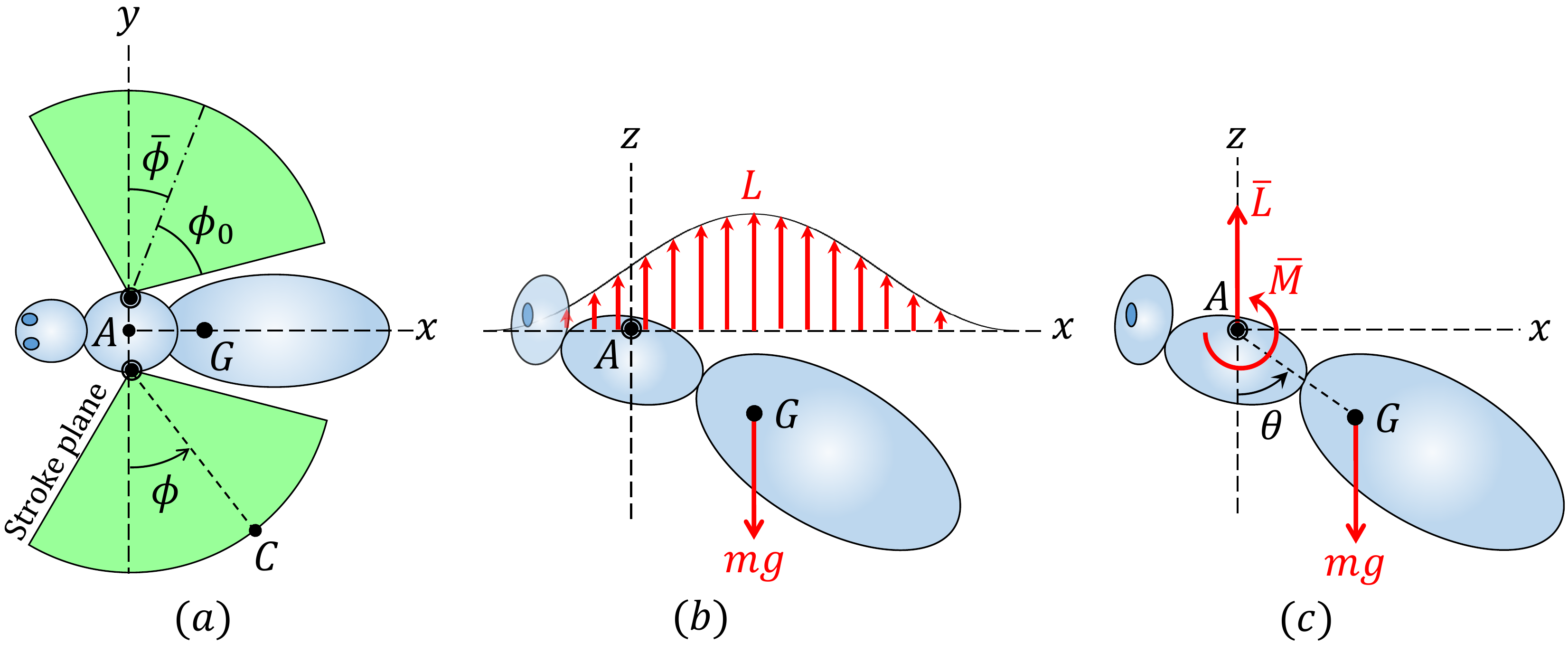}
\caption{a) Top view of the asymmetric flapping~\eqref{eqn:FlapKin1}, b) the nonzero-mean forces acting on the body of a hovering insect with the asymmetric flapping kinematics~\eqref{eqn:FlapKin1} during one period, and c) the equivalent average force-couple system of the lift.}
\label{fig:Forces}
\end{figure}

Figure~\ref{fig:Forces} shows the asymmetric flapping~\eqref{eqn:FlapKin1}, the nonzero-mean forces during hover, i.e., lift and weight, and the equivalent average force-couple system acting on the body of a hovering insect due to asymmetric flapping during one period. As shown in that figure, the lift may be replaced by an equivalent force-couple system, on average, with a force $\bar{L}$ and couple $\bar{M}$.

%%%%%%%%%%%%%%%%%%%%%%%%%%%  FIGURE  %%%%%%%%%%%%%%%%%%%%%%%%%%%%%%%
%%%%%%%%%%%%%%%%%%%%%%%%%%%%%%%%%%%%%%%%%%%%%%%%%%%%%%%%%%%%%%%%%%%%

\begin{figure}[thpb]
\centering
\includegraphics[width=6 in]{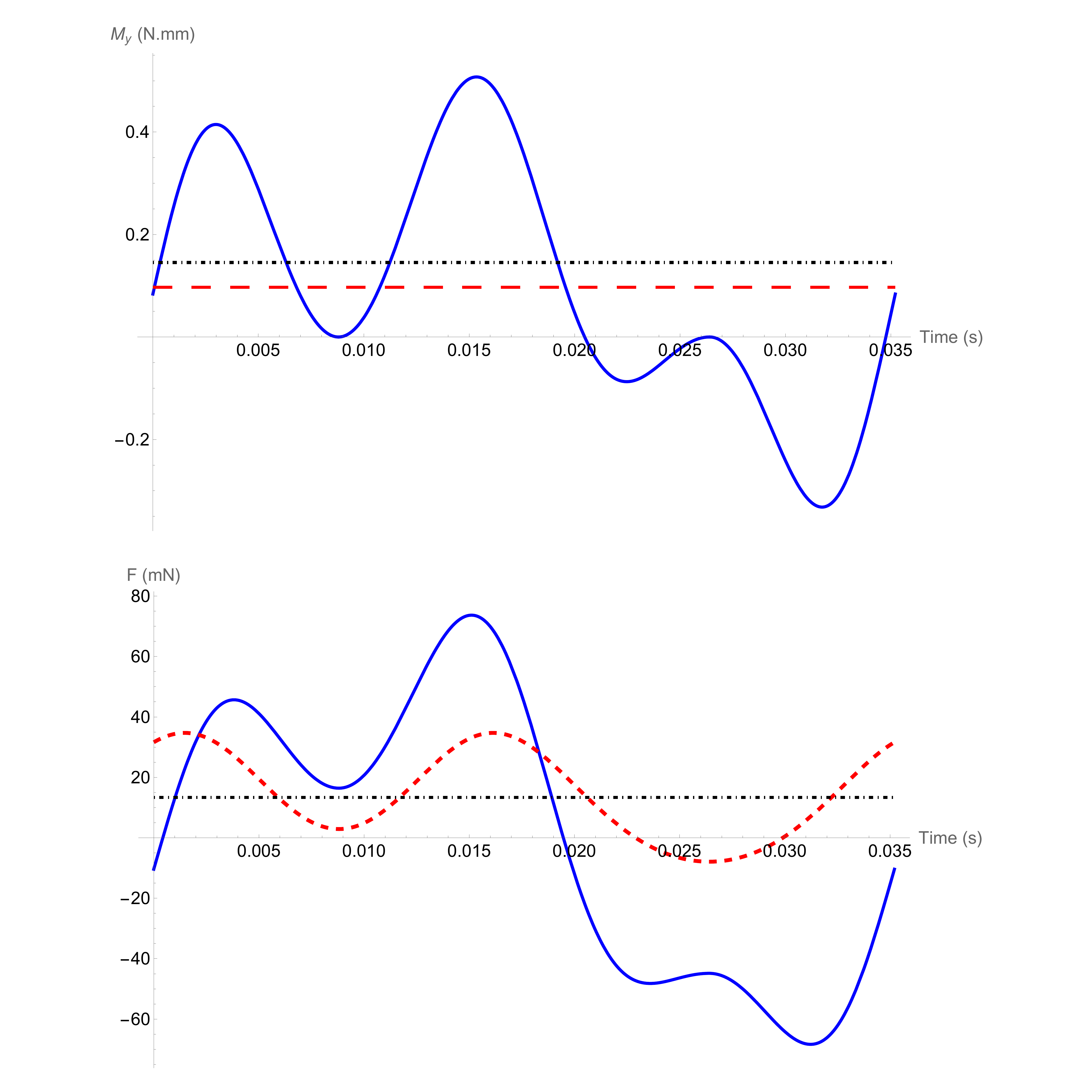}
\caption{Total moment and forces acting on the body of a hovering hawkmoth over one flapping period. Top: solid-blue is the moment $M_y$, dashed-red is the average moment $\bar{M}$, and dot-dashed-black is $mgd$ for comparison. Bottom: solid-blue is $F_x$, dashed-red is $F_z$, and dot dashed-black is $mg$ for comparison.}
\label{fig:HMFM}
\end{figure}

\begin{figure}[thpb]
\centering
\includegraphics[width=8 in]{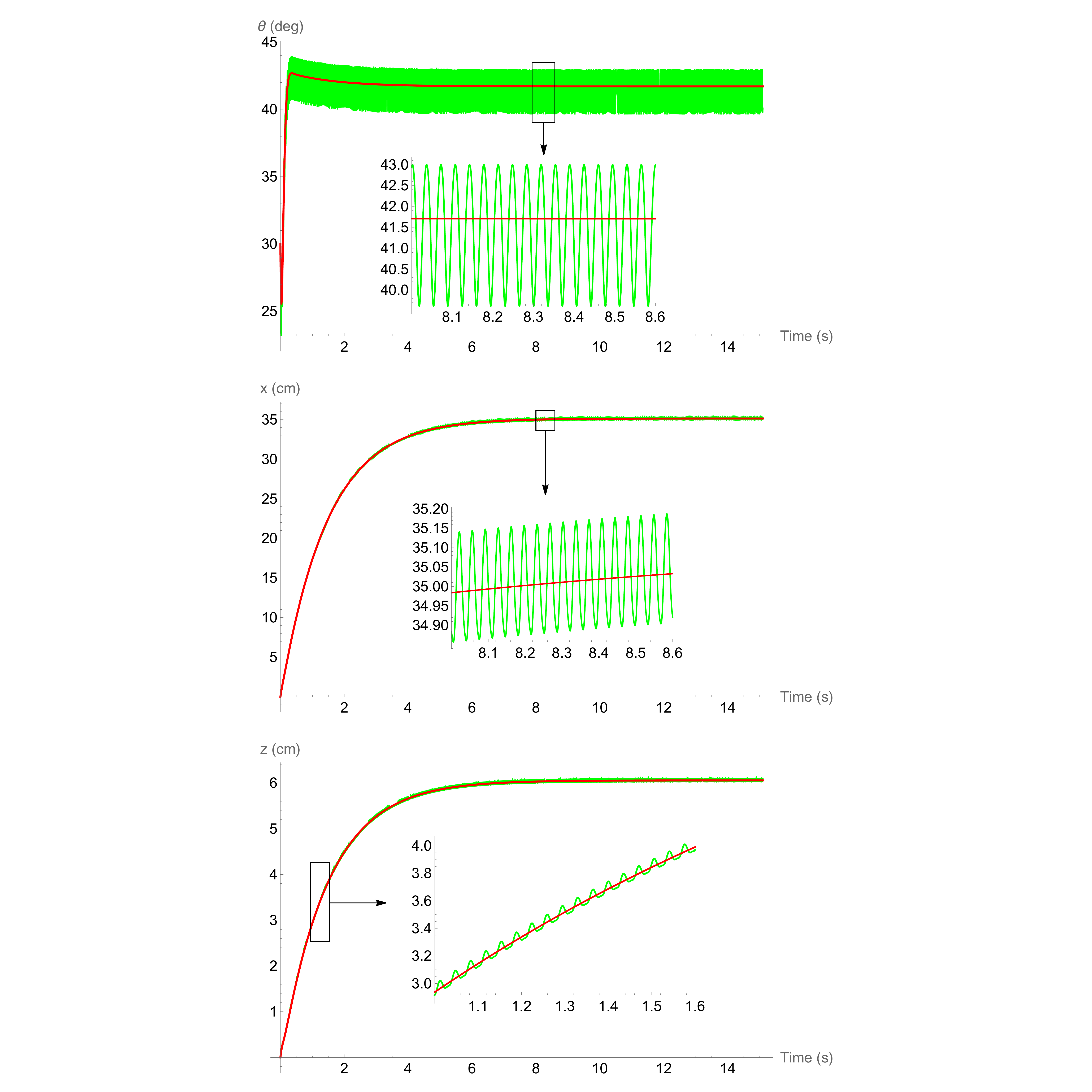}
\caption{Time history of the hovering hawkmoth. Solid-green: original system, solid-red: averaged dynamics.}
\label{fig:HM}
\end{figure}

Figures~\ref{fig:HMFM} and~\ref{fig:HM} show the total moment $M_y$ and forces $F_x$ and $F_z$ and the simulation results for hovering of the hawkmoth with its morphological parameters presented in Tables~\ref{tab:Table1}, \ref{tab:Table2}, and~\ref{tab:Table4}. The initial conditions of the simulations presented in Figure~\ref{fig:HM} are $x(0)=z(0)=0$, $\theta(0)=30^{\circ}$, and zero initial velocities. The stability of the equilibrium point of the averaged dynamics and the corresponding periodic orbit of the original time-periodic system can be clearly seen in Figure~\ref{fig:HM}.

%%%%%%%%%%%%%%%%%%%%%%%%%%%  FIGURE  %%%%%%%%%%%%%%%%%%%%%%%%%%%%%%%
%%%%%%%%%%%%%%%%%%%%%%%%%%%%%%%%%%%%%%%%%%%%%%%%%%%%%%%%%%%%%%%%%%%%

%%%%%%%%%%%%%%%%%%%%%%%%%%%  FIGURE  %%%%%%%%%%%%%%%%%%%%%%%%%%%%%%%
%%%%%%%%%%%%%%%%%%%%%%%%%%%%%%%%%%%%%%%%%%%%%%%%%%%%%%%%%%%%%%%%%%%%

%%%%%%%%%%%%%%%%%%%%%%%%%%%  FIGURE  %%%%%%%%%%%%%%%%%%%%%%%%%%%%%%%
%\begin{figure}[thpb]
%\centering
%\includegraphics[width=5 in]{BodyStrokeAngles.pdf}
%\caption{The body angle ($\theta$) and flapping angle ($\phi$) of the hawkmoth during steady-%state hovering flight. Dashed-blue: body angle, solid-red: flapping angle.}
%\label{fig:BSA}
%\end{figure}
%%%%%%%%%%%%%%%%%%%%%%%%%%%%%%%%%%%%%%%%%%%%%%%%%%%%%%%%%%%%%%%%%%%%

\begin{figure}[thpb]
\centering
\includegraphics[width=5 in]{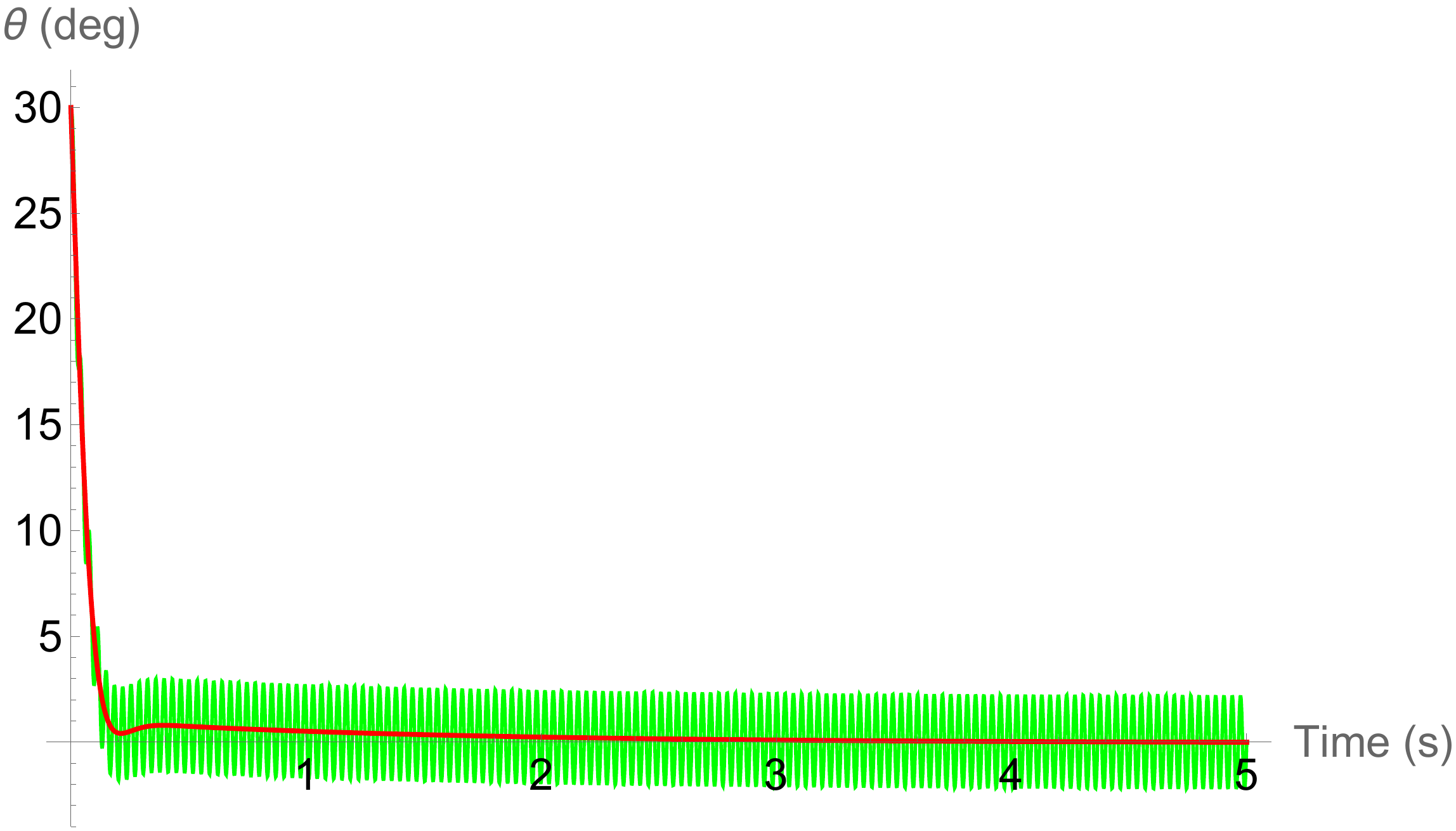}
\caption{Time history of the hovering hawkmoth with symmetric flapping, $\bar{\phi}=0$. Solid-green: original system, solid-red: averaged dynamics.}
\label{fig:HMSym}
\end{figure}

As mentioned in Section~\ref{sec:InsectIF}, the results suggest that without the flapping asymmetry, the aerodynamic and inertial forces are not enough to put the body of a hovering insect in a non-vertical orientation, as seen in real world. To show this, the results of numerical simulation of the dynamics of a hovering hawkmoth with symmetric flapping kinematics, that is, $\bar{\phi}=0$, is presented in Figure~\ref{fig:HMSym}. The parameters and initial conditions are the same as used to generate the results in Figure~\ref{fig:HM}, except that the mean stroke angle is considered to be zero ($\bar{\phi}=0$). It is evident that the body cannot be stabilized anymore and moves to an almost downright orientation.

%%%%%%%%%%%%%%%%%%%%%%%%%%%  FIGURE  %%%%%%%%%%%%%%%%%%%%%%%%%%%%%%%
%%%%%%%%%%%%%%%%%%%%%%%%%%%%%%%%%%%%%%%%%%%%%%%%%%%%%%%%%%%%%%%%%%%%

\begin{figure}[thpb]
\centering
\includegraphics[width=3.0 in]{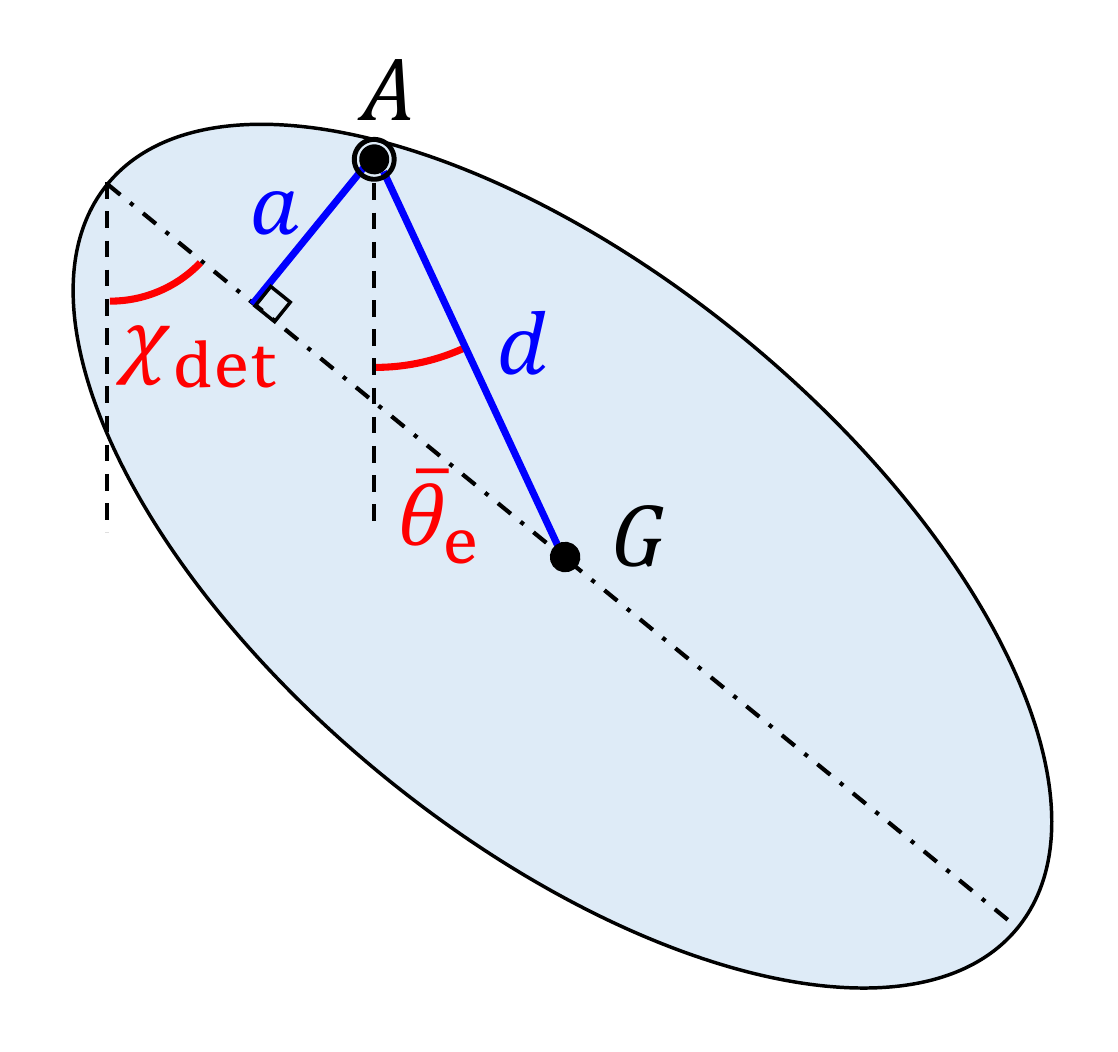}
\caption{The determined body angle $\chi_{\rm det}$ and the equilibrium orientation $\bar{\theta}_{\rm e}$.}
\label{fig:Angles}
\end{figure}

Table~\ref{tab:Table4} also presents the average equilibrium orientation $\bar{\theta}_{\rm e}$ of the body during hover determined using~\eqref{eqn:BodyAngle} for the five insect species. The underlined data could not be found in the literature and are estimated. The orientation angle $\theta$ considered in this paper is the angle of the line $AG$, and not the body itself, with the vertical (see Figure~\ref{fig:Insect}). This angle is smaller than the real body angle. As an estimation of the body angle determined using the averaged dynamics, one may use the geometry shown in Figure~\ref{fig:Angles} where the insect body is shown as an ellipse. To determine the body angle of each insect species, one may use the determined equilibrium orientation $\bar{\theta}_{\rm e}$ and the two lengths $d$ and $a$ shown in Figure~\ref{fig:Angles}, and determine the body angle $\chi_{\rm det}$ using
\beq
\chi_{\rm det} = \bar{\theta}_{\rm e} + \sin^{-1}\left( \frac{a}{d}\right)
\label{eqn:AproxBodyAngle}
\eeq
The values of the length $a$ and the determined body angle $\chi_{\rm det}$ are presented in Table~\ref{tab:Table4}. Besides the determined body angle $\chi_{\rm det}$, Table~\ref{tab:Table4} also presents the real body angle $\chi_{\rm obs}$ of the five insect species observed during experiments and reported in the literature used for the morphological data and also in \cite{EllingtonP3.PTRS.1984,Willmott&Ellington.JEB.1997,VanceEtAl.PBZ.2014}. The determined body angles using the averaged dynamics for the five insect species show good agreement with the observed body angles.

%%%%%%%%%%%%%%%%%%%%%%%%%%%  FIGURE  %%%%%%%%%%%%%%%%%%%%%%%%%%%%%%%
%%%%%%%%%%%%%%%%%%%%%%%%%%%%%%%%%%%%%%%%%%%%%%%%%%%%%%%%%%%%%%%%%%%%

%%%%%%%%%%%%%%%%%%%%%%%%%%%%%%%   TABLE   %%%%%%%%%%%%%%%%%%%%%%%%%%%%%%%%%
\begin{table}[ht]
    \caption{The damping coefficients $c \: ({\rm N.s/m})$ and $c_t \: ({\rm N.m.s/rad})$, the eigenvalues $\lambda_i$ of the linearized averaged dynamics, the length $a$, the equilibrium orientation $\bar{\theta}_{\rm e}$, the determined body angle $\chi_{\rm det}$, and the observed body angle $\chi_{\rm obs}$ of the five insect species.}
    \begin{center}
    \begin{tabular}{|c|c|c|c|c|c|c|c|}
    \cline{2-8}
    \multicolumn{1}{c|}{} & \rule{0pt}{12pt} $c$ & $c_t$ & $\lambda_i$ & $a \: ({\rm mm})$ & $\bar{\theta}_{\rm e} \: (^{\circ})$ & $\chi_{\rm det} \: (^{\circ})$ & $\chi_{\rm obs} \: (^{\circ})$ \\
    \hline
    HM & $10^{-3}$ & $10^{-5}$ & $-17.2 \pm 8.18i, \: -0.687, \: -0.686$ & \underline{3.5} & 42 & 60 & 54  \\

    HF & $10^{-5}$ & $10^{-8}$ & $-191.7, \: -3.39, \: -0.927, \: -0.926$ & 0.3 & 69 & 89 & 78 \\
    
    DF & $10^{-5}$ & $10^{-7}$ & $-66.2, \: -18.3, \: -0.111, \: -0.111$ & 1.1 & 25 & 62 & 52 \\
    
    HB & $10^{-5}$ & $10^{-7}$ & $-22.7 \pm 27.0i, \: -0.097, \: -0.097$ & 1.3 & 34 & 57 & 50 \\
    
    BB & $10^{-4}$ & $10^{-6}$ & $-461.8, \: -5.45, \: -0.569, \: -0.568$ & 1.4 & 36 & 57 & 44 \\
    \hline
    \end{tabular}
    \label{tab:Table4}
\end{center}
\end{table}
%%%%%%%%%%%%%%%%%%%%%%%%%%%%%%%%%%%%%%%%%%%%%%%%%%%%%%%%%%%%%%%%%%%%%%%%%%%

The results from the stability analysis of the pitch motion of hovering insects presented in this section show that the main parameter determining the pitch stability and the average body angle of hovering insects is the average moment $\bar{M}$ generated by asymmetric flapping which counteracts the moment due to the insect weight about the wing hinges. The nonzero average aerodynamic moment $\bar{M}$ is the result of an \emph{asymmetric} flapping kinematics, such as~\eqref{eqn:FlapKin1}. The wing and body damping causes the equilibrium body orientation determined from~\eqref{eqn:BodyAngle} to be an open-loop asymptotically stable equilibrium of the averaged dynamics. The wing inertial forces, which as shown, are larger than aerodynamic forces in amplitude, also play a minor role in the equilibrium orientation of the body during hover. Surprisingly, as mentioned in Section~\ref{sec:Introduction}, in many papers discussing the pitch stability of insects during hovering flight, the main parameters determining the stability of the pitch motion during hover, that is, the nonzero mean stroke angle $\bar{\phi}$ (the flapping asymmetry), the distance between the wing hinge axis and the body center of mass, and the wing inertial forces, are neglected.

%%%%%%%%%%%%%%%%%%%%%%%%%%%   Subsection   %%%%%%%%%%%%%%%%%%%%%%%%%%%%%%%%%%%
\subsection{Experimental results}
\label{sec:Experiment}
\begin{figure}[thpb]
\centering
\includegraphics[width=3.4 in]{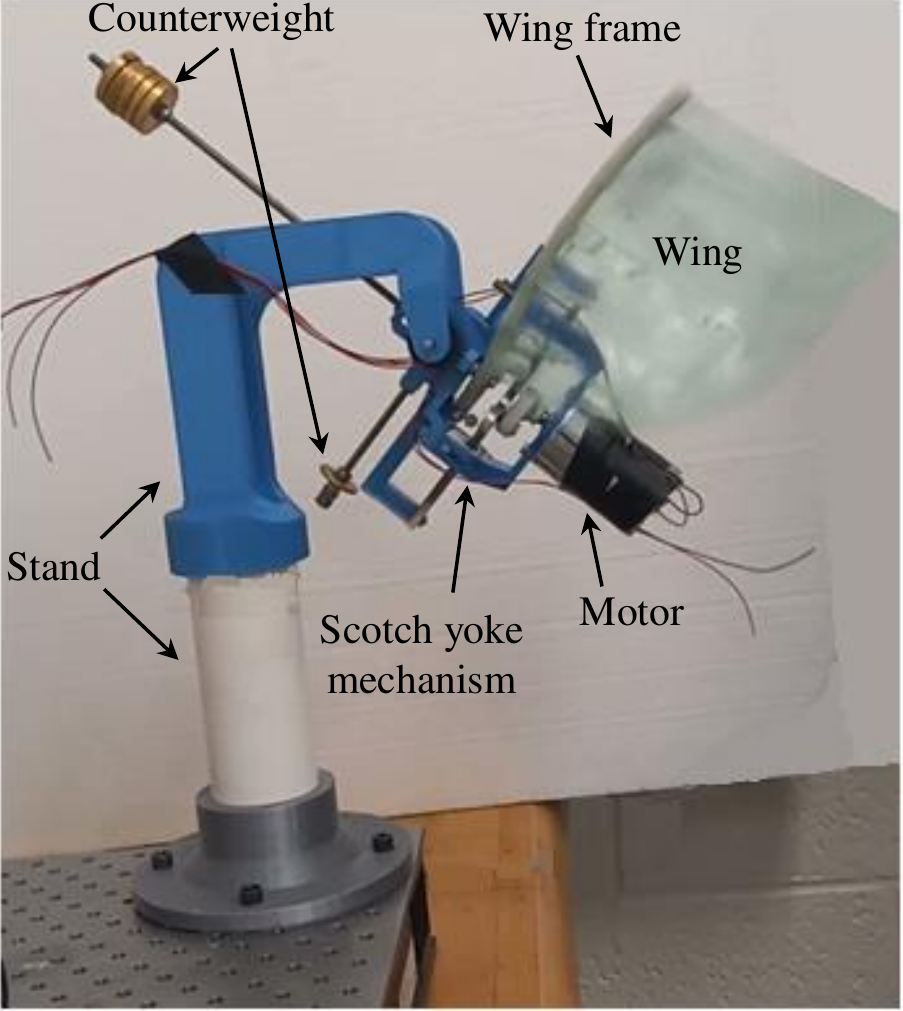}
\caption{The flapping wing device.}
\label{fig:FWD}
\end{figure}

%%%%%%%%%%%%%%%%%%%%%%%%%%%%%%%%%%%%%%%%%%%%%%%%%%%%%%%%%%%%%%%%%%%%%%%%%%%
To demonstrate the effect of the mean stroke angle in stability of the pitch dynamics of hovering insects, we designed a flapping wing device consisting of a 1-DOF main body with two wings attached to it, as shown in Figure~\ref{fig:FWD}. The body, which represents the insect body, is free to rotate about a fixed horizontal shaft. Each wing consists of a light flexible membrane attached to a rigid arm (frame). The wings are driven back and forth by a DC motor and a Scotch yoke mechanism \cite{Norton.2012}. The flapping mechanism can be adjusted to flap with either a zero or nonzero mean stroke angle, that is, symmetric or asymmetric flapping kinematics. Using carefully selected counterweights, the system is made slightly heavier on one side. Therefore, the center of mass of the device is located outside of its axis of rotation, that is, the fixed horizontal shaft. Without flapping, the system is a 1-DOF pendulum that is stable in its vertical orientation. The goal of the experiments is to show that the body can be stabilized in a non-vertical orientation while flapping with a \emph{nonzero} mean stroke angle (asymmetric flapping). However, it cannot be stabilized using flapping with a \emph{zero} mean stroke angle (symmetric flapping). In other words, the goal is to experimentally show that the aerodynamic moment generated by asymmetric flapping can stabilize the pitch dynamics of the main body in a non-vertical orientation.

%%%%%%%%%%%%%%%%%%%%%%%%%%%  FIGURE  %%%%%%%%%%%%%%%%%%%%%%%%%%%%%%%
%%%%%%%%%%%%%%%%%%%%%%%%%%%%%%%%%%%%%%%%%%%%%%%%%%%%%%%%%%%%%%%%%%%%

As shown in the video available on \url{https://youtu.be/PWSiN-znN7k}, in the first experiment we used a symmetric flapping kinematics and tried to stabilize the body in a non-vertical orientation. However, symmetric flapping was not able to stabilize the pitch dynamics and the body remained close to the vertical orientation on average. 

In the second experiment we used an asymmetric flapping kinematics that was able to stabilize the pitch dynamics in a non-vertical orientation. Using a higher flapping frequency, we were able to stabilize the body in an almost horizontal orientation. To show that the pitch stability is caused by the aerodynamic moment generated by asymmetric flapping, and not the vibrational effects of the wing inertial forces, in the third experiment the membranes are removed from the wing arm (frame). The membranes are light and their total mass and inertial forces are negligible compared to the mass and inertial forces of the wing frame and other reciprocal parts. The membranes generate almost the entire aerodynamic forces during flapping, and by removing them the aerodynamic effects vanish. As is evident in the video of the third experiment, though we used the same asymmetric flapping kinematics as in the second experiment, the pitch dynamics cannot be stabilized in a non-vertical orientation without the aerodynamic moment. Although in the experimental device the main body is a 1-DOF pendulum, compared with the 3-DOF body of a hovering insect, the experiments clearly show that the asymmetric flapping kinematics plays the most important role on the stability of the pitch dynamics of hovering insects.

%%%%%%%%%%%%%%%%%%%%%%%%%%%   Section   %%%%%%%%%%%%%%%%%%%%%%%%%%%%%%%%%%%
\section{Conclusions}
\label{sec:Conclusion}
%%%%%%%%%%%%%%%%%%%%%%%%%%%%%%%%%%%%%%%%%%%%%%%%%%%%%%%%%%%%%%%%%%%%%%%%%%%
First this paper discussed the stability and open-loop vibrational control of a 3-DOF pendulum with vibrational force and moment inputs. It was shown that the pendulum requires a minimum force magnitude to be stabilized in a non-vertical orientation. The results were used for stability analysis of the pitch motion of hovering insects. The body of a hovering insect may be considered as a 3-DOF pendulum with vibrational inputs consisting of aerodynamic and inertial forces and moments. Using numerical values, it was shown that, in general, the inertial forces due to flapping of the wings are larger than the aerodynamic forces during hovering flight. However, the inertial and aerodynamic forces and moments together may not be large enough to \emph{vibrationally} stabilize the insect body in a non-vertical orientation. Instead, the pitch stability of insect bodies during hover may be due to two counteracting moments, the moment of the body weight about the wing hinges and the nonzero-mean aerodynamic moment generated due to an \emph{asymmetric} flapping kinematics. Using numerical values for five insect species, it was shown that their stable body angle during hover predicted by the analysis presented in this paper agrees with the body angles of real hovering insects observed in experiments. The results of this paper suggest that the two main parameters determining the body angle and pitch stability of hovering insects, namely, the distance between the body center of mass and the wing hinge axis and the nonzero mean stroke angle (flapping asymmetry) should not be ignored in the pitch dynamics and stability analysis of hovering insects. Experiments with a 1-DOF flapping device confirmed these results. Also, it was shown that the inertial forces due to flapping of the wings effect the body angle and should be considered in the dynamic analysis of hovering insects.

%%%%%%%%%%%%%%%%%%%%%%%%%%%   Bibliography   %%%%%%%%%%%%%%%%%%%%%%%%%%%%%%
\bibliography{References}
\bibliographystyle{unsrt}
%%%%%%%%%%%%%%%%%%%%%%%%%%%%%%%%%%%%%%%%%%%%%%%%%%%%%%%%%%%%%%%%%%%%%%%%%%%

\end{document}